\documentclass[epj,final,amsmath,amssymb]{svjour}

\usepackage{graphicx,amsmath}
\usepackage{latexsym}
\usepackage{graphicx}
\usepackage{subfigure}
\begin{document}

\title{Google matrix of business process management}
\author{M. Abel\inst{1} \and D.L. Shepelyansky\inst{2}}
\institute{Department of Physics and Astronomy, Potsdam University,
Karl-Liebknecht-Str 24, D-14476, Potsdam-Golm, Germany 
\and 
Laboratoire de Physique Th\'eorique du CNRS, IRSAMC, 
Universit\'e de Toulouse, UPS, F-31062 Toulouse, France}
\titlerunning{Google matrix of business management}
\authorrunning{M.Abel and D.L.Shepelyansky}


\date{Received: September 14, 2010}

\abstract{
Development of efficient business process models and determination of their
characteristic properties are  subject of intense interdisciplinary research. 
Here, we consider a business process model as a directed graph. Its nodes correspond 
to the units identified by the modeler and the link direction indicates 
the causal dependencies between units. 
It is of primary interest to obtain 
the stationary flow on such a directed graph, which corresponds 
to the steady-state of a firm during the business process. 
Following the ideas developed recently for the World Wide Web,
we construct the Google matrix for our business process model and analyze
its spectral properties. The importance of nodes is characterized
by Page\-Rank and recently proposed CheiRank and 2DRank, respectively.
The results show that this two-dimensional ranking gives a significant
information about the influence and communication properties
of business model units.
We argue that the Google matrix method, described here, 
provides a  new efficient tool helping companies to make 
their decisions on how to evolve in the exceedingly dynamic global market. 
} 
%
\maketitle

\section{Introduction}

Business process models are dynamical systems that describe the interdependencies
of functional units, or components, on a micro- or macroeconomic level. 
They depict the way a company works and eventually makes money 
with the strategy it uses. The efficiency of a model is 
primarily determined by the help a model can give for strategic decisions, 
e.g. if a reorientation of products or 
marketing is needed due to changes in the market or 
opportunities because of technological developments 
(see e.g. \cite{Sterman-01,bpm} and Refs. therein).

The building of a business model is a complicated task, 
because all important units 
in the company value production must be identified and 
properly linked {\em at a certain level} of modeling. 
This involves a cancellation of non-important unit, which might be even
harder. What modelers do further is a qualitative identification 
if a unit positively or negatively stimulates a linked one 
(amplification or damping, respectively). This yields a directed graph, 
where the units of the model are linked and the direction reflects causality. 
The next step towards quantitative modeling is 
the prescription of a functional dependence of the units, 
which is basically a very heuristic procedure. Clearly, the functions 
have to be nonlinear, because a growth to plus/minus infinity is not allowed, 
so typical functions are of sigmoid-type, on the other hand minimal models 
are of predator-prey type, well known from biology. This reflects 
the modern point of view of a company as a quasi-organic, dynamical system.

In this work we introduce and analyze 
the Google Business Process Model (GBPM) of a real consulting 
company \cite{Grasl-08} whose
major product is of intellectual nature. The detailed description
of the original dynamical model can be found in \cite{Grasl-08}
and thus we do not present it here. The model describes a dynamical workflow 
propagation (see e.g. \cite{leymann1,leymann2}) which is simulated by
certain dynamical equations.

In our approach we trace parallels and similarities between
the directed graph of this model and the Google matrix approach
used for the ranking of the World Wide Web (WWW) \cite{brin,meyer,avrach3}.
Thus, we investigate only  the model graph and 
do not enter the subject of dynamical simulations, 
because we want to reveal the underlying structure of the stationary state 
of the model without using the quite heuristic functional dependencies 
which need to be further supported by statistical analysis and measurement. 
This is not to say that the latter is a wrong approach, however 
the determination of the stationary density by the application of 
the Page\-Rank algorithm for the Google matrix, 
which is a variant of Frobenius--Perron operator \cite{meyer}, 
is a very powerful and well--established technique which gives fundamental
results on the network without solving the dynamical equations and using 
a vast study of parameter variations.

Indeed, the construction of the Google matrix for the WWW
and the determination of the stationary probability distribution over
WWW network via the Page\-Rank algorithm has been proposed by Brin and Page \cite{brin}
and by now it became a powerful tool for classification of the WWW nodes
(see e.g. \cite{meyer,avrach3,donato,upfal}). The approach based on the Google matrix 
construction for a directed network is rather general and finds applications for
various types of networks including 
university WWW networks \cite{giraud},
Ulam networks of dynamical maps
\cite{zhirov2010ul,ermann2010ul}, brain neural networks \cite{brain},
procedure call network of Linux kernel \cite{aliklinux,leolinux},
hyperlink network of Wikipedia articles \cite{zhirovwiki}.
Page\-Rank finds also applications in blog analysis \cite{fortunato1},
citation network of Phys. Rev. \cite{redner,fortunato2},
and food flow network between species in ecosystems \cite{food}.

In this work we extend this approach to the network of business management.
How is the model built? Basically, one has identified major 
components of the company, 
which are refined in their dynamics in respective subcomponents. 
By construction, the model is hierarchical, but links between components 
can be set according to the needs of the modeler. 
We only mention here the  components
and the nodes in the top component: {\em managers, consultants,...};
subcomponents are: 
{\em top, consultants, products, 
proposals, customers, ....} . 
The full list of nodes and links between them 
are given in Appendix.
Depending on the business process, one of the nodes is the most important one, 
followed by others. 
This is the value of our method: 
we {\em identify without any bias the most important components} 
in a model. This provides an extremely 
helpful information. If these components are not 
the ones wished by the shareholders or management, respectively, 
the model has to be changed and adapted. 
Since the computation is not very costly 
this gives a tool to simulate small changes, 
e.g. by linking different nodes, and 
studying their effect on the business process model.
We consider the GBPM as a first
 step in the application of the Google matrix analysis
to the business process management. Next steps should extend this approach 
and take into account actual workflow between nodes
inside a company\cite{leymann1,leymann2} .

\begin{figure}
 \begin{center}
\includegraphics[clip=true,angle=270,width=0.7\textwidth,angle=90]{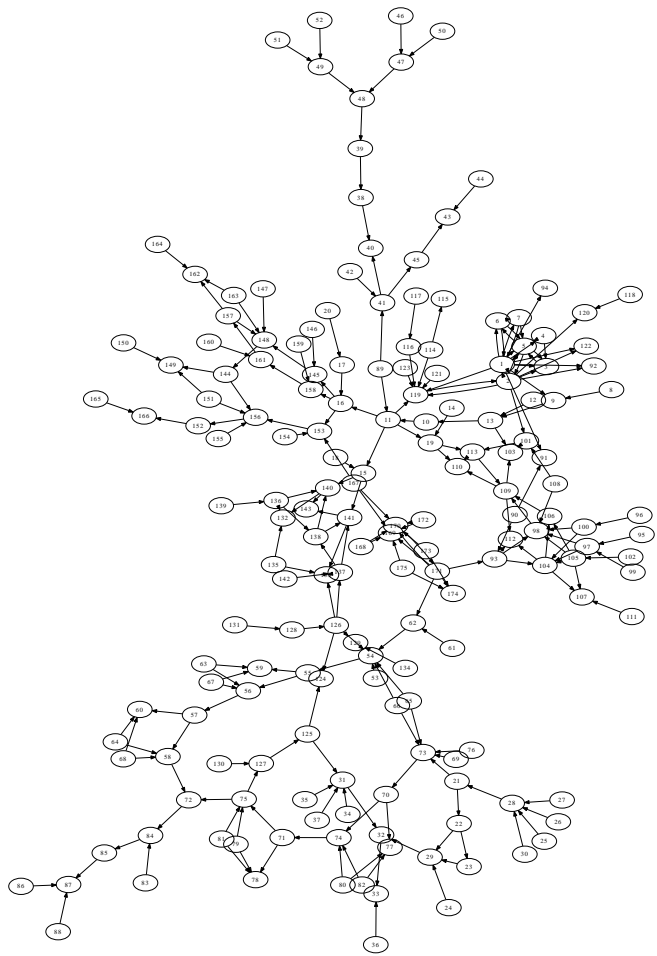}
\vglue -0.1cm
\caption{ 
Google Business Process Model with links taken from \cite{Grasl-08}. The network is 
structured into several subgraphs reflecting the functionality of the model.
The names (or meaning) of the nodes and links between them
are listed in the Appendix.
\label{fig1}
}
\end{center}
\end{figure}

Our network is small in comparison of typical applications of Google Matrix, 
like the WWW \cite{donato,upfal}, Linux kernel network \cite{aliklinux,leolinux}
or Wikipedia network \cite{zhirovwiki}.
It consists of 175 nodes only and is graphically displayed in 
Fig.~\ref{fig1}. This size is comparable
with the one of food network in ecosystems \cite{food}.
Our purpose is an elementary study of 
the network properties using the spectral characteristics of the Google matrix,
Page\-Rank and recently introduced CheiRank and 2DRank such that the order $10^2$ is sufficient; 
the latter ranking algorithms are explained in detail below, 
Most big business models are proprietary (for understandable reasons), 
and an application of the Google matrix method is straightforward.

\begin{figure}
 \begin{center}
\includegraphics[clip=true,width=8.0cm]{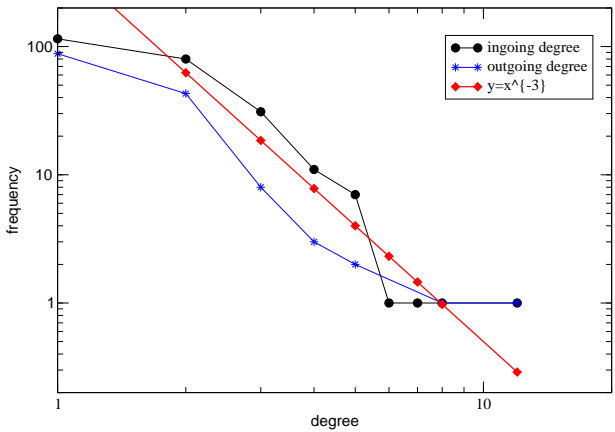}
\vglue -0.1cm
\caption{ (Color online)   
Distribution of ingoing (black points) and 
outgoing (blue points) links. An approximate global 
power law scaling with the exponent 3 is shown by the straight red line.
\label{fig2}}
\end{center}
\end{figure}

Let us have a look on the network in terms of connectivity: 
the distribution of ingoing and outgoing links is shown in 
Fig.~\ref{fig2}. Of course, with only one decade 
available it is useless to 
try to identify exact scaling behaviour; 
nevertheless the global distribution 
is compatible with  power law scaling 
$f(d)\sim d^{-\nu}$ at $\nu \approx 3$.
The exponent $\nu=3$ is not so far from the exponent
$\nu=2.1$ and $2.7$ found for the WWW
for ingoing and outgoing link distributions respectively
\cite{donato,upfal}. 
It will be interesting to investigate the generic 
scaling of business models in the future for 
networks of larger size.

\section{Method}

The Google matrix $G$ underlies the determination of Page\-Rank \cite{brin},
which is a tool used by virtually every Internet user 
when issuing an Internet search for some keywords. This approach gives 
a powerful and general way to analyze networks. For the construction of the
Google matrix 
we use the procedure described in \cite{brin,meyer}:
\begin{equation}
 G_{ij} =  \alpha S_{ij} + (1-\alpha)/N\;,
\label{eq1}
\end{equation}
where $S_{ij}$ is the normalized adjacency  matrix of the graph.
The elements of the adjacency matrix are zero (if there is no link)
or one  (if there is a link). Due to the normalization the sum 
of all  elements inside one column is equal to unity.
Columns with zeros only are replaced by $(1/N,\dots,1/N)$, 
with $N$ being the network size. 
Because it is a full stochastic matrix of a Markov chain,
the matrix  ${\bf S}$
has N eigenvalues $\lambda_i$, $i=1,\dots,N$
which are generally complex. In agreement
with the Perron-Frobenius theorem (see e.g. \cite{meyer})
the largest eigenvalue is $\lambda_1=1$. 
The damping parameter $\alpha$ denotes the possibility for a random surfer 
on the graph to jump to any other node. Its effect is to  bound away the eigenvalues  
with absolute value smaller than one: $|\lambda_i| \leq \alpha<1$
for $i>1$. 
A typical value, used as well for the WWW search, is $\alpha=0.85$, 
however this choice can be varied without essential 
impact on the results presented below. 
The  right eigenvectors, $\psi_i$, are defined by
${\bf G} \psi_i=\lambda_i \psi_i$, cf. \cite{meyer,giraud}.
The Page\-Rank vector is the one with $\lambda=1$, and  
since ${\bf G}$ is a Frobenius-Perron operator, the corresponding 
right eigenvector, $\psi_1 = (P(1),\dots,P(N))^T$  
gives the stationary probability density $P(i)$ 
that a random surfer is found at site $i$ with $\sum_i P(i)=1$. 
Once it is found, the nodes are sorted according to 
decreasing $P(i)$, the node rank in this index, 
$K(i)$ corresponds  to its relevance.

Other eigenvalues correspond to non-stationary, decaying modes.
They are of transient nature and may play an important role in 
non-stationary considerations, because they may live for a long time before
dying out. This is, however, not the focus of this work.

\section{CheiRank versus PageRank}
\begin{figure}
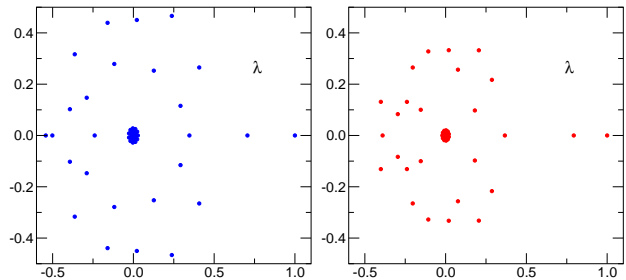

\includegraphics[clip=true,width=4.0cm]{fig3a.eps}
\includegraphics[clip=true,width=4.0cm]{fig3b.eps}
\vglue -0.1cm
\caption{ (Color online)   Distributions of eigenvalues $\lambda$ of 
the Google matrix at $\alpha=0.85$ in the complex plane 
for matrix ${\bf G}$ (left panel) and matrix ${\bf G^*}$
with inverted link directions (right panel).
\label{fig3}}
\end{figure}

In a nutshell, the procedure uses the idea that a node is not only relevant 
if it is highly linked. One has also to take into account the relevance of the nodes 
pointing to it. Since this is an iterative procedure, 
the Page\-Rank vector can be easily computed by the so--called power-iteration 
using consecutive multiplication of initially random vector 
on the Google matrix  \cite{meyer}. 
Of course, this vector is the most important one, because it represents 
the stationary distribution on the graph. 
The relaxation process to the steady-state given by the Page\-Rank
is affected by the eigenmodes with $|\lambda|$ close  to $\alpha$.
It is known that for the WWW there are many eigenvalues
which are close or even equal to $\alpha$ (see e.g. \cite{meyer,giraud}).
The spectrum of the Google matrix ${\bf G}$ of the GBPM is shown in the left panel of
Fig.~\ref{fig3}. The eigenvalue next after $\lambda=1$
is $\lambda_2=0.706$ and other eigenvalues have $|\lambda| < 0.52$.
There are only about 14\% of eigenvalues with $|\lambda| > 0.1$
that gives an indication on a possibility of appearance of the fractal Weyl law
for such type of networks of larger size $N$ in analogy with the
Linux kernel network analyzed in \cite{aliklinux,leolinux}.
The spectrum of the Google matrix ${\bf G^*}$, obtained from the 
network with the inversed direction of links, is shown in
the right panel of Fig.~\ref{fig3}, its characteristics are
similar to those of matrix ${\bf G}$.

\begin{figure}
\includegraphics[clip=true,width=0.35\textwidth,angle=270]{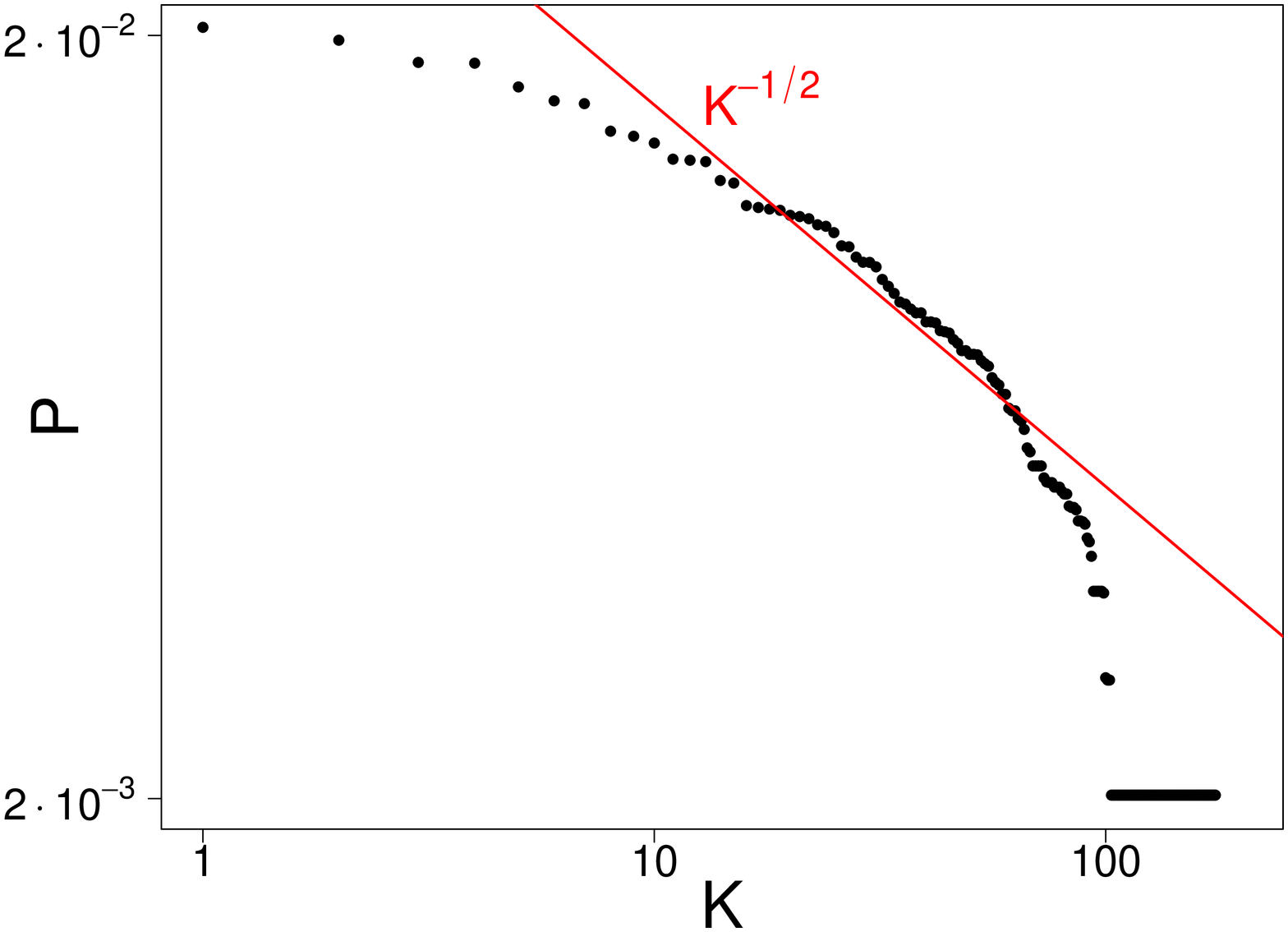}
\includegraphics[clip=true,width=0.35\textwidth,angle=270]{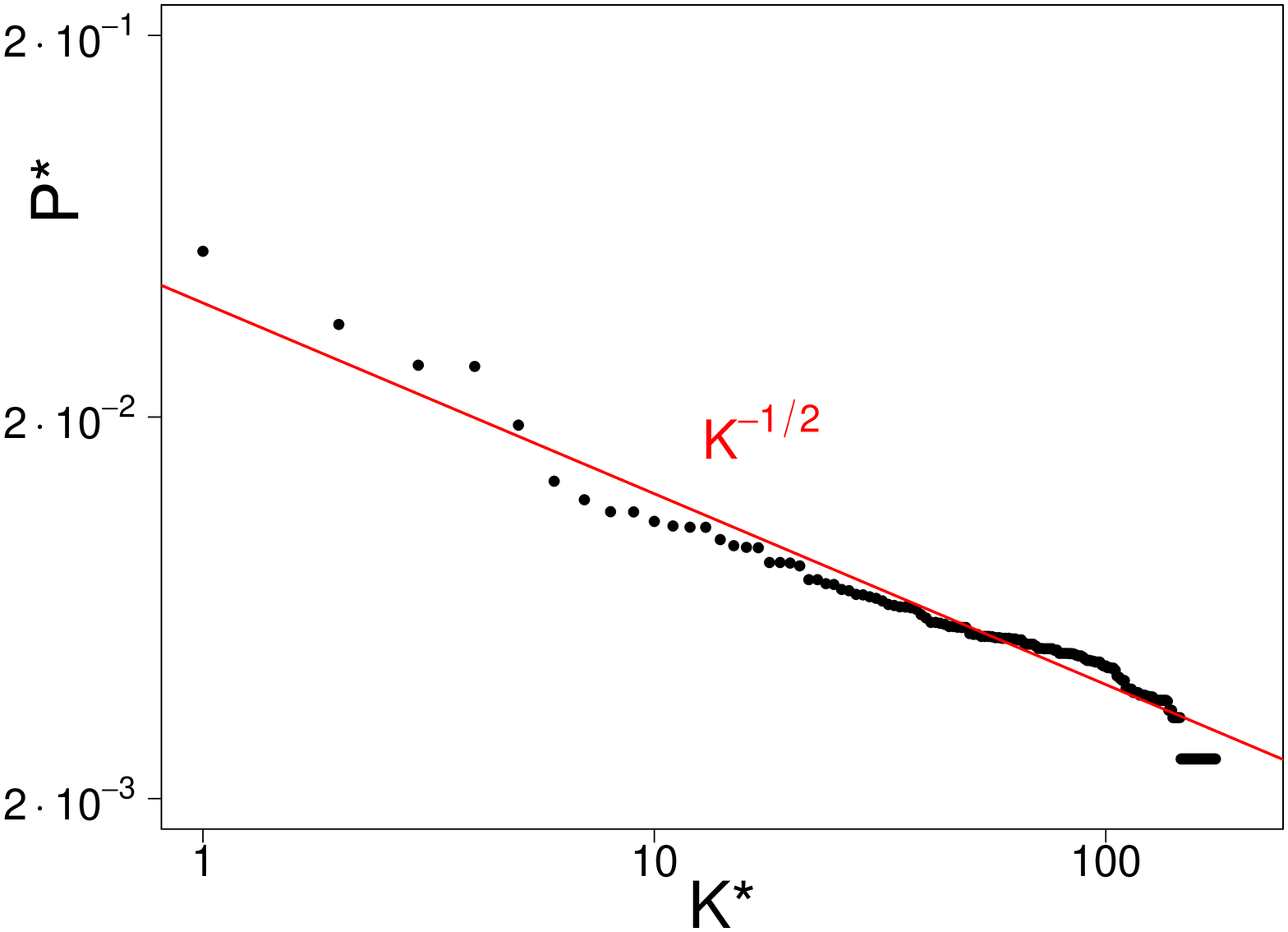}
\vglue -0.1cm
\caption{Top panel: probability of Page\-Rank vector $P(i)$ 
as a function of PagRank $K(i)$ in log-log scale.
Bottom panel: probability of CheiRank vector  $P^*(i)$ in log-log scale. 
The  straight lines show the approximate power law dependence
with the slope $1/(\nu-1)=1/2$, corresponding the the average slope 
$\nu=3$ shown in Fig.~\ref{fig2}.
\label{fig4}}
\end{figure}

The Page\-Rank probability $P(i)$ for our business model 
is shown in Fig.~\ref{fig4} (top panel) 
as a function of rank $K(i)$. Surprisingly, there is no dominant node, 
which means that this 
company is quite democratic - in terms of relevance. The first five nodes are:
{\it Identified Contact Loss (33), Identified Contacts (32), Projects (5), 
Consultants (2),  Delivery Project Completion (87)}.
The numbers in brackets denote the node indices, cf. the Appendix.
Managers (node index 1) do not appear before rank 18. 
This is quite surprising, since the management is expected to be at least 
among top ten positions. 
How can one understand that behaviour?
The management plays typically the role of coordinating projects 
and {\it keeping all together}, which means that they decide 
which points are most important and have many outgoing links
related to orders given to others. However, the Page\-Rank is proportional in average
to the number of ingoing links \cite{meyer}. 
This implies the management units are not most important
according to the Page\-Rank since they do not have 
a large number of ingoing links
(not many units give order to managers).
In the considered model of a consulting company 
the most relevant units are the customers, or contacts. 
Without them, no business is made, especially for 
consulting. The first two ranks can be explained by this. 
The following ranks are {\it Projects} and {\it Consultants}. Of course,
without good projects and correspondingly good workers 
the firm will die, so this is of vital relevance. Rank 5 again involves projects, 
this time their delivery. This means that in this model the way  
the projects are completed is given a high importance. This might not be necessarily 
true in all cases, however for the model of the firm under consideration it is. 
We recognize that in this view the result makes perfect sense: 
customers, products and consultants are 
the most relevant units in the model of a consulting firm. 
Such a firm can only survive when its consultants are top level 
and its products are alike - and if there are customers. 
The management is responsible {\it only} to get the firm running well. 
This result may be surprising, but reveals the power of the method.  
This means as well that the most attention  for refinement of the model should 
be put on the top nodes given above.
Nevertheless,  one expects that the management plays somehow a very influential role.

It is interesting to note that a similar situation takes place for 
the procedure call network of the Linux kernel as it was shown in \cite{aliklinux}.
Indeed, for this network the Page\-Rank gives at the top procedures which
are often pointed on but which are not so much important for the code
functionality. Thus it was proposed \cite{aliklinux}
to characterize the network also by the Page\-Rank of the Google matrix 
obtained from the network with inversed link directions.
The rank $P^*(i)$ of this inversed matrix ${\bf G^*}$, named as
the CheiRank \cite{zhirovwiki}, places on first positions rather influential code
procedures. Hence, it is natural to use the CheiRank also for our
model of business process management.

And indeed, using the CheiRank, introduced in \cite{aliklinux} 
we obtain an adequate result. 
It corresponds to the stationary distribution, $P^*(i)$, of the inverted flow, 
or the information returned from the nodes to their precedent ones. 
Thus, it describes the influence or communication ranking of the nodes. 
Again, the eigenvector 
with the eigenvalue 1 is computed and sorted according to the magnitude of the entries. 
This yields a new rank, $K^*(i)$, the mentioned CheiRank. 
The result of the computation of $P^*(i)$ vs. $K^*(i)$ 
is displayed in Fig.~\ref{fig4}. 
(bottom panel). Here, we can also give a tentative scaling 
$P^*(i) \sim K^{1/(\nu-1)}$ which must be compared and verified, 
respectively, with other business models of larger size. 
While the distribution of $P(i) \sim  K^{1/(\nu-1)}$ is proportional to the distribution of
ingoing links, the distribution of $P^*(i)$ is proportional
to the distribution of outgoing links
(see e.g. \cite{meyer,avrach3,giraud,aliklinux}).
Due to a small size of our network
we do not try to use different values of $\nu$
for ingoing and outgoing links and for $P$ and $P^*$ respectively.
According to the CheiRank the top nodes are:
{\it Principals (1), Projects (5), Consultants (2), Customers (6), Contacts (7)}. 
The management now has clearly 
first position in the ranking which is fully logical, 
since any management decision {\em influences} the whole company, 
while the management is not necessarily 
the most {\em important} component, as explained above.
\begin{figure}
\includegraphics[width=0.38\textwidth,angle=270]{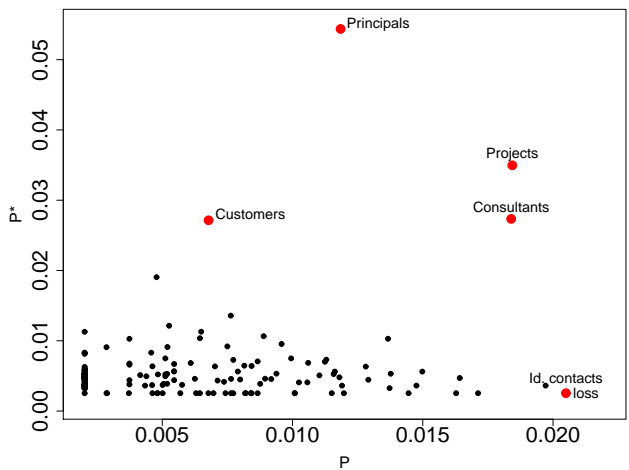}
\vglue -0.1cm
\caption{ (Color online)   Distribution of nodes in the plane 
of probabilities of Page\-Rank 
$P(i)$ and CheiRank $P^*(i)$. The marked nodes illustrate the first four nodes in Chei rank (Principals, Projects, Consultants, Customers), and the top node in PageRank (Identified Contacts Loss)
\label{fig5}}
\end{figure}

\begin{figure}
\includegraphics[width=0.38\textwidth,angle=270]{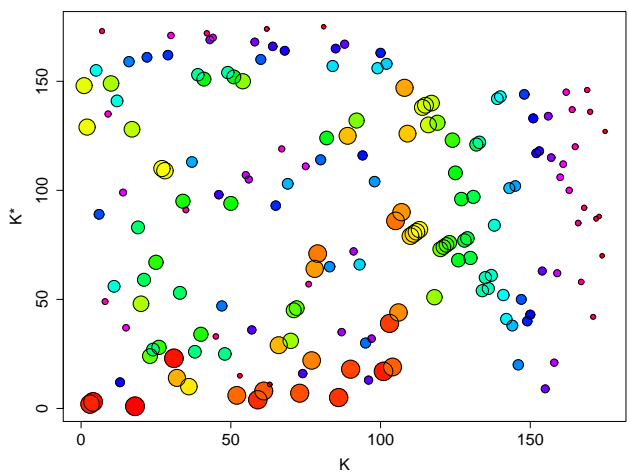}
\vglue -0.1cm
\caption{ (Color online)   Distribution of nodes in the plane 
of Page\-Rank  $K$ and CheiRank $K^*$, size of circles and their color
is proportional to their listing node index with large radius (red color)
for small index and small radius (blue-ros\'e) for large index.
\label{fig6}}
\end{figure}

\begin{figure}
\includegraphics[width=0.38\textwidth,angle=270]{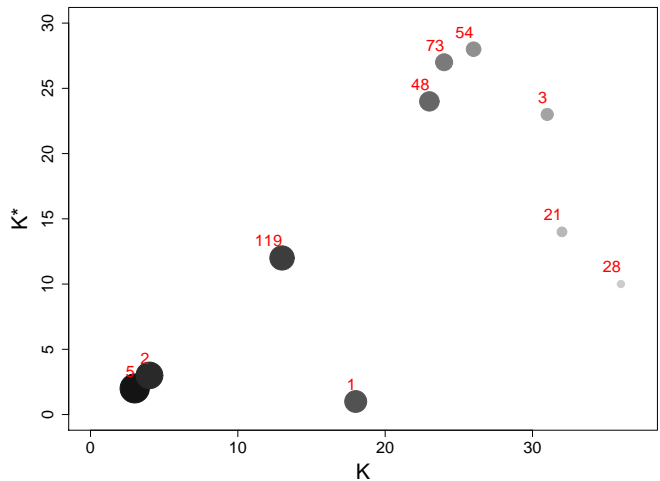}
\vglue -0.1cm
\caption{ Zoom of the distribution of nodes in the plane 
of Page\-Rank $K$ and CheiRank $K^*$ in the region of small $K, K^*$
values. Numbers near circles give the listing node index,
grayness is proportional to 2DRank $K_2$ with black for minimum
and  light gray for maximum $K_2$ (see Appendix).
\label{fig7}}
\end{figure}

Following \cite{aliklinux} we also use the joint distribution of nodes 
in the  plane of probabilities
$(P(i),P^*(i))$ of Page\-Rank and CheiRank shown in Fig.~\ref{fig5}.
That way, we see both ranks at once and can decide which emphasis to put, 
defining importance in a new way. In this sense, the most important nodes are 
indicated in Fig.~\ref{fig5}. The distribution of all nodes in the plane of 
Page\-Rank and CheiRank $(K,K^*)$ is shown in Fig.~\ref{fig6}.
In the plane $(K,K^*)$ the most important nodes are those with the
smallest values of $K$ and $K^*$. The zoom of this region of the plane
is shown in Fig.~\ref{fig7}.

Of course, nodes might be both relevant  (well-known) and 
influential (communicative). This can be characterized by the correlator $\kappa$
between Page\-Rank and CheiRank which is defined as
\begin{equation}
\kappa = N \sum_i P(i) P^*(i)-1 \; .
\label{eq2}
\end{equation}
For the WWW university networks \cite{aliklinux} and 
Wikipedia network \cite{zhirovwiki} it was found that the correlator
is rather large with $\kappa \approx 4$  while for the Linux kernel network
one has very small correlator $\kappa \approx -0.05 \ll 1$.
For the GBPM we have $\kappa=0.164$ showing that there is practically
no correlations between nodes with large number of 
outgoing and ingoing links. Thus the GBPM network
has more similarities with the Linux kernel network
in contrast to the WWW and Wikipedia networks which are
characterized by high correlations between 
nodes which are highly known (high Page\-Rank) and
highly communicative (high CheiRank).

With the appearance of CheiRank all nodes are now distributed in a
two-dimensional plane (see Figs.~\ref{fig5},\ref{fig6},\ref{fig7}).
How can one combine both rankings in a way to find  nodes which 
are  both very relevant and influential?
There are many ways to find such a single-valued 
one-dimensional ranking which combines $K$ and $K^*$: one can think of the
distance $(K^2+{K^*}^2)$, or the absolute value, or some other combination of $K$ and $K^*$. 
Since $P(K)$ and $P^*(K^*)$ are monotonic functions the plane $(K,K^*)$
is mapped into $(P,P^*)$ plane in a unique way. 

\begin{figure}
 \begin{center}
\includegraphics[clip=true,width=6.8cm,angle=270]{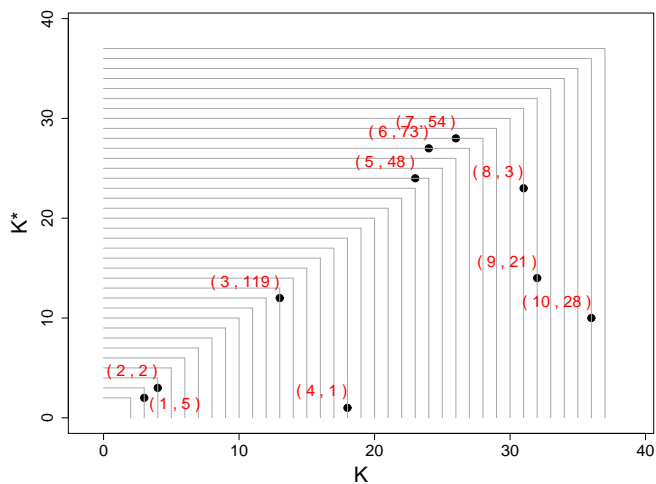}
 \caption{Illustration of the 2DRank algorithm to find rank $K_2$ 
which combines Page\-Rank $K$ and CheiRank $K^*$. 
Specific nodes are drawn in the $(K,K^*)$ plane 
when crawling through the squares, indicated by the grey lines, 
from small to large $(K,K^*)$ the nodes are labeled by $K_2$;
numbers in brackets $(K2(i),i)$ give 
the value of found 2DRank  $K_2$ and 
the values of listing node index $i$. 
One recognizes that at most 2 nodes can be found on a square edge, and some
edges might be empty.
}
\label{fig8}
\end{center}
\end{figure}

A convenient way to order all nodes of the two-dimen\-sional plane on a
one-dimensional line was proposed in \cite{zhirovwiki}
for Wikipedia articles being named 2DRank $K_2$.
This rank is described by the algorithm presented below;
it is dubbed 2DRank $K_2$, since it combines the two ranks
discussed above.
Remember that a ranking is basically a list of pairs (rank and nodes index), in our case $K2,i$, or
simply $K_2(i)$.
By $K_2$, we also use this ordering of nodes by the   following,  quite 
intuitive criterion: 
we look progressively if a point $(K,K^*)$ lies on the square
$j \times j$, where $j$ is a running index starting at 1. 
Since the ordering is unique, there are only two possibilities for this to
occur: either $K=j$ or $K^*=j$. It may happen, that neither $K$ nor $K^*$ 
lies on the square, 
then one increases $j$ by one and compares again with $(K,K^*)$.
The initial $K_2$ list is empty.
E.g. if there is no point with $K=1$ and
$K^*=1$, then the first square $1\times 1$ 
has no point on it and the next square $2\times 2$ is considered. 
The algorithm works by setting $j=1$, 
then we look if $K=j$, if yes, $i(K,K^*)$ is determined and added to the list
$K_2(i)$ whose own running index is increased; 
then we apply this procedure to $K^*$: if $K^*=j$, the node index 
$i(K,K^*)$ is determined and added to the list $K_2(i)$.
Since there are no more points to check, 
we step from $j$ to $j+1$. 
The algorithm is finished if all nodes $i$ have been visited.
We can deliberately choose if we first look for $K$ or $K^*$ (we have chosen
first $K$). 
The procedure is illustrated for the first ten nodes in K2 ranking in Fig.~\ref{fig8}.

According to this 2DRank algorithm we find for the first five nodes in 2DRank $K_2$:  
{\it Projects (5), Consultants (2), 
Hire Rate (119), Principals (1), Required Delivery Proposal Effort (48)}. 
The principals are still not the most relevant node, but obviously this
ranking gives a quite balanced characterization of the business process
management under consideration.

Top 30 nodes ordered according to Page\-Rank, CheiRank and 2DRank
are given in Appendix. Ranking of all nodes is available
at the website \cite{gbpmpage}.

\section{Discussion}
We have presented a powerful method which
quantitatively describes the business process management 
in terms of the Google matrix, its eigenvectors and eigenvalues. 
The application of the method yields the stationary distribution on 
the directed graph which describes the business process of a concrete company
in the frame of our GBPM. Our results show that the importance and influence
of the units of business process are well characterized by 
two-dimensional ranking in the plane defined by Page\-Rank and CheiRank.
These ranks show that certain units (e.g. {\it Contacts})
perform important tasks being highlighted by  Page\-Rank, while other units
(e.g. {\it Principals })  realize influential communication processes
highlighted by  CheiRank. Thus the two-dimensional ranking 
described here establishes
a broad and detailed characterization of main operational units
of business process management.
In contrast to the WWW university networks and Wikipedia network,
the network of GBPM has rather small correlation between
top units of Page\-Rank and CheiRank that stresses a clear separation between
communication and  realization tasks of business process.
In this respect the GBPM network is more similar to the 
procedure call network of Linux kernel which also has small
correlation between these two ranks.

Of course, the approach developed here is in its initial stage 
and more advanced business process modeling will need 
weighted graphs with subgraphs for the flows of work, information, money, products, etc.
These generalizations are straightforward and can be constructed at next more
advanced stage.  A study of changes in the model is quick and straightforward, 
such that systematic studies of future activities of a company are now
feasible without 
sometimes very heuristic equations which can be used at a final modeling
stage. But now one is relieved from the task to 
determine fine--tune parameters and equations each time a model is changed. 
We expect these results to have significant impact in econometry for 
the evaluation of small, middle-size and large-scale models
of business process management. The application to macro-economy is straightforward, 
and global flows might be characterized by the GBPM procedure.

\section*{Acknowledgements}
We acknowledge fruitful discussion with O. Grasl who kindly provided 
his model \cite{Grasl-08} to us and explained 
the basics of business process modeling.

\paragraph{Appendix}

\paragraph{List of Nodes}
(node number is followed by its name and comma):

1	Principals,
2	Consultants,
3	Value,
4	Products,
5	Projects,
6	Customers,
7	Contacts,
8	Heads Of Branch,
9	Total Principals,
10	Maximum Principal Proposal Effort,
11	Maximum Principal Hiring Effort,
12	Average Principal Work Effort,
13	Maximum Principal Work Effort,
14	Maximum Project Time Share,
15	Maximum Contact Maintenance Effort,
16	Maximum Product Effort,
17	Contact Maintenance Effort,
18	Maximum Contact Maintenace Time Share,
19	Maximum Principal Project Effort,
20      Contacting Effort,
21	Qualified Contacts,
22	Required Contact Maintenance Effort,
23	Qualified Contact Maintenance Effort,
24	Qualified Contact Lifetime,
25	Maximum Qualified Contacts,
26	Minimum Qualification Duration,
27	Qualification Fraction,
28	Contact Qualification Rate,
29	Qualified Contact Loss,
30	Maximum Qualification Rate,
31	Contact Identification,
32	Identified Contacts,
33	Identified Contact Loss,
34	New Customer Contact Potential,
35	Identificaton Duration,
36	Identified Contact Lifetime,
37	Identification Fraction,
38  Delivery Proposal Effort,
39  New Delivery Proposal Effort,
40  Delivery Proposal Writing Effort,
41  Principal Delivery Proposal Effort,
42  Delivery Proposal Effort Share,
43  Delivery Proposal Closing Rate,
44  Delivery Proposal Writing Rate,
45  Minimum Duration Per Delivery Proposal,
46  Delivery Project Effort, 
47  Effort Per Delivery Proposal,
48  Required Delivery Proposal Effort,
49  Delivery Lead Success Rate,
50  Delivery Proposal Effort Fraction,
51  First Time Delivery Lead Success,
52  Repeat Delivery Lead Success,
53   Repeat Delivery Lead Fraction,
54   Repeat Delivery Lead Generation,
55   Repeat Delivery Leads,
56   Repeat Delivery Lead Success,
57   Repeat Delivery Proposals,
58   Repeat Delivery Proposal Success,
59   Repeat Delivery Lead Loss,
60   Repeat Delivery Proposal Loss,
61   Delivery Project Effort,
62   Customer Delivery Lead Generation Duration,
63   Delivery Lead Closing Duration,
64   Delivery Proposal Closing Rate,
65   Lead Generation Pressure,
66   Effect Of Delivery Project Per Principal,
67   Repeat Delivery Lead Success Fraction,
68   Repeat Delivery Proposal Success Fraction,
69   First Time Delivery Lead Generation Duration,
70   First Time Delivery Leads,
71   First Time Delivery Proposals,
72   Delivery Projects Won,
73   First Time Delivery Lead Generation,
74   First Time Delivery Lead Success,
75   First Time Delivery Proposal Success,
76   First Time Delivery Lead Fraction,
77   First Time Delivery LeadLoss,
78   First Time Delivery Proposal Loss,
79   Delivery Proposal Closing Rate,
80   Delivery Lead Closing Duration,
81   First Time Delivery Proposal Success Fraction,
82   First Time Delivery Lead Success Fraction,
83   Average Time To Delivery Project Start,
84   Delivery Project Start,
85   Active Delivery Projects,
86   Delivery Project Effort,
87   Delivery Project Completion,
88   Delivery Project Completion Rate,
89  Principal Proposal Effort,
90  Active Delivery Projects,
91  Delivery Project Per Principal,
92  Total Consulting Staff,
93  Delivery Projects Staff Needed,
94  Consultants Needed,
95  Active Consulting Projects,
96  Active Solution Projects,
97  Consulting Projects Staff Needed,
98   Project Work Rate Needed,
99   Consulting Project Leverage,
100  Solution Projects Staff Needed,
101  Maximum Consultant Work Effort,
102  Solution Project Leverage,
103  Utilization Percentage,
104  Total Project Staff Needed,
105  Solution Projects Staff Needed,
106  Solution Project Delivery Rate,
107  Delivery Project Completion Rate,
108  Average Work Rate,
109  Actual Project Delivery Rate,
110  Principal Project Effort,
111  Delivery Projects Staff Needed,
112  Consulting Project Delivery Rate,
113  Maximum Work Rate,
114 Hiring Effort Per Hire,
115 Hiring Effort,
116 Consultant Target,
117 Annual Consultant Growth Target Percentage,
118 Fluctuation Rate,
119 Hire Rate,
120 Fluctuation,
121 Maximum Leverage,
122 Leverage
123 Average Hiring Duration,   
124 Total Customers,
125 New Customers,
126 Mature Customers,
127 Customer Acquisition,
128 Customer Maturing,
129 Customer Attrition,
130 Customer Project Conversion,
131 Maturing Duration,
132 New Customer Loss,
133 Mature Customer Loss,
134 Customer Lifetime,
135 Customer ErosionTime,
136 Required New Customer Maintenance Effort,
137 Required Mature Customer Maintenance Effort,
138 New Customer Contact Maintenance Effort Share,
139 New Customer Maintenance Effort Per Customer,
140 New Customer Contact Maintenance Effort,
141 Mature Customer Contact Maintenance Effort,
142 Mature Customer Maintenance Effort Per Customer,
143 Customer Maintenance Effort,
144 Marketable Product,
145 Product Marketing Effort,
146 Product Marketing Effort Percentage,
147 Required Product Marketing Effort,
148 Product Marketing Rate,
149 Marketing Reject,
150 Development Reject Duration,
151 Development Reject Fraction,
152 Standardised Product,
153 Product Standardisation Effort,
154 Product Standardisation Effort Percentage,
155 Required Product Standardisation Effort,
156 Product Standardization Rate,
157 Innovation Product,
158 Poduct Innovation Effort,
159 Product Innovation Effort Percentage,
160 Required Product Innovation Effort,
161 Product Innovation Rate,
162 Innovation Reject,
163 Innovation Reject Fraction,
164 Innovation Reject Duration,
165 Product Lifetime,
166 Product Obsolescence Rate,
167 Time To Standardisation,
168 Leverage Adjustment Time,
169 Leverage Loss,
170 Leverage Win,
171 Project Leverage,
172 Time To Standardization Excellence,
173 Maximum Project Leverage,
174 Project Leverage Percentage,
175 Minimum Project Leverage.

\paragraph{List of Links}
(node number marked by dot is followed by numbers of nodes
on which it points to, last node number or blanc if empty is marked by comma):

1.   2 3 4 5 6 7 9 91 92 94 119 122,
2.   1 3 5 92 101 119 120 122,
3.   5,
4.   5 3,
5.   1 2 3 6,
6.   5 7 1,
7.   5 1,
8.   9,
9.   13,
10.  11,
11. 19 15 16 119,
12.  13,
13.  10 103,
14.  19,
15.  140 141,	
16.  145 153 158, 	
17.  16,
18.  15,
19.  110 113,	
20.	,
21.  22 73,
22.  23 29,
23.  29,
24.  29,
25.  28,
26.  28,
27.  28,
28.  21,
29.  32,
30.  28,
31.  32,
32.  33,
33.    ,
34.  31,
35.  31,
36.  33,
37.  31,
38.  40,
39.  38,
40.    ,
41.  40 45,
42.  41,
43.    ,
44.  43,
45.  43,
46.  47,
47.  48,
48.  39,
49.  48,
50.  47,
51.  49,
52.  49,
53.  54,
54.  55,
55.  56 59,
56.  57,
57.  58 60,
58.  72,
59.    ,
60.    ,
61.  62,
62.  54,
63.  56 59,
64.  58 60, 
65.  54 73,
66.  54 73,
67.  56 59,
68.  58 60,
69.  73,
70.  74 77,
71.  75 78,
72.  84,
73.  70,
74.  71,
75.  72 127,
76.  73,
77.    ,
78.    ,
79.  75 78,
80.  74 77,
81.  75 78,
82.  74 77,
83.  84,
84.  85,
85.  87,
86.  87,
87.    ,
88.  87,
89. 41,
90.  91 93,
91.    ,
92.    ,
93.  98 104 112,
94.    ,
95.  97,
96.  100,
97.  104 98,
98.  109,
99.  97,
100. 98 104,
101. 103 113,
102. 105,
103.   ,
104. 106 107 112,
105. 98 104 106 107, 
106. 109,
107.     ,
108. 98 101,
109. 103 110 112,
110.     ,
111.  107,
112.     ,
113. 109 110,
114. 115 119,
115.        ,
116. 119,
117. 116,
118. 120,
119. 2,
120.  ,
121. 119,
122.    ,
123. 119,
124.    ,
125. 31 124, 
126. 54 124 129 133 137, 
127. 125,
128. 126,
129.    ,
130. 127,
131. 128,
132.    ,
133.    ,
134. 129,
135. 132 133,
136. 132 138 140,
137. 133 138 141, 
138. 140 141,
139. 136,
140. 132 143,
141. 133 143,
142. 137,
143.    ,
144. 149 156,
145. 148,
146. 145,
147. 148,
148. 144,
149.    ,
150. 149,
151. 149 156,
152. 166,
153. 156,
154. 153,
155. 156,
156. 152,
157. 148 162,
158. 161,
159. 158,
160. 161,
161. 157,
162.    ,
163. 148 162, 
164. 162,
165. 166,
166.    ,
167. 153 169 170,
168. 169 170,
169.       ,
170. 171,
171. 62 93 169 170 174, 
172. 169 170,
173. 170 174,
174.        ,
175. 169 174,

\paragraph{Page\-Rank top 30 nodes:}
1 Identified Contact Loss,
2 Identified Contacts,
3 Projects,
4 Consultants,
5 Delivery Project Completion,
6 Actual Project Delivery Rate,
7 Product Obsolescence Rate,
8 Product Standardization Rate,
9 Standardised Product,
10 Delivery Proposal Writing Effort,
11 Delivery Project Start,
12 Active Delivery Projects,
13 Hire Rate,
14 Marketable Product,
15 Product Marketing Rate,
16 Utilization Percentage,
17 Delivery Proposal Effort,
18 Principals,
19 Delivery Projects Won,
20 New Delivery Proposal Effort,
21 First Time Delivery Leads,
22 Principal Project Effort,
23 Required Delivery Proposal Effort,
24 First Time Delivery Lead Generation,
25 Repeat Delivery Leads,
26 Repeat Delivery Lead Generation,
27 Contact Identification,
28 Qualified Contact Loss,
29 Consulting Project Delivery Rate,
30 Marketing Reject.

\paragraph{CheiRank top 30 nodes:}
1 Principals,
2 Projects,
3 Consultants,
4 Customers,
5 Contacts,
6 Maximum Principal Work Effort,
7 Maximum Principal Proposal Effort,
8 Maximum Principal Hiring Effort,
9 Maturing Duration,
10 Contact Qualification Rate,
11 Leverage Win,
12 Hire Rate,
13 Customer Maturing,
14 Qualified Contacts,
15 Project Leverage,
16 Mature Customers,
17 Products,
18 Total Principals,
19 Average Principal Work Effort,
20 Solution Project Leverage,
21 New Customer Maintenance Effort Per Customer,
22 Maximum Product Effort,
23 Value,
24 Required Delivery Proposal Effort,
25 First Time Delivery Proposal Success,
26 First Time Delivery Lead Success,
27 First Time Delivery Lead Generation,
28 Repeat Delivery Lead Generation,
29 Required Contact Maintenance Effort,
30 Solution Projects Staff Needed .

\paragraph{2DRank top 30 nodes:}
1 Projects,
2 Consultants,
3 HireRate,
4 Principals,
5 RequiredDelivery Proposal Effort,
6 First Time Delivery Lead Generation,
7 Repeat Delivery Lead Generation,
8 Value,
9 Qualified Contacts,
10 Contact Qualification Rate,
11 Product Marketing Rate,
12 First Time Delivery Lead Success,
13 Repeat Delivery Lead Success,
14 Product Innovation Rate,
15 Total Project Staff Needed,
16 First Time Delivery Proposal Success,
17 New Delivery Proposal Effort,
18 Product Standardization Rate,
19 Maximum Principal Work Effort,
20 Project Leverage,
21 First Time Delivery Proposals,
22 Delivery Project Start,
23 Customer Acquisition,
24 Customers,
25 First Time Delivery Leads,
26 Maximum Principal Hiring Effort,
27 Leverage Win,
28 Required Contact Maintenance Effort,
29 Repeat Delivery Leads,
30 Principal Delivery Proposal Effort.

\end{document}